\documentclass[prd,reprint,onecolumn,notitlepage,nofootinbib,showpacs,superscriptaddress]{revtex4-1}

\usepackage{graphicx} 
\usepackage{hyperref}
\usepackage{amsfonts}
\usepackage{amsmath,amssymb,latexsym}
\usepackage{bm}
\usepackage{color}
\usepackage{xcolor}
\usepackage[caption=false]{subfig} 
\usepackage{float}
\usepackage{url}
\usepackage{multirow}
\usepackage{tablefootnote}
\usepackage{capt-of}
\usepackage{textcomp}

\begin{document}


\title{The Effects of a Nuclear Disturbed Environment on a Quantum Free Space Optical Link}


\author{David~A.~Hooper}
\email{hooperda@ornl.gov}
\affiliation{Detonation Forensics and Response Group of the Nuclear Nonproliferation Division, Oak Ridge National Laboratory, Oak Ridge, Tennessee 37831, USA}

\author{Brandon~A.~Wilson}
\affiliation{Detonation Forensics and Response Group of the Nuclear Nonproliferation Division, Oak Ridge National Laboratory, Oak Ridge, Tennessee 37831, USA}

\author{Alexander~Miloshevsky}
\affiliation{Quantum Information Science Group, Computational Sciences and Engineering Division, Oak Ridge National Laboratory, Oak Ridge, Tennessee 37831, USA}
\affiliation{Department of Mechanical, Aerospace, and Biomedical Engineering, University of Tennessee, Knoxville, Tennessee 37996, USA}

\author{Brian~P.~Williams}
\affiliation{Quantum Information Science Group, Computational Sciences and Engineering Division, Oak Ridge National Laboratory, Oak Ridge, Tennessee 37831, USA}

\author{Nicholas~A.~Peters}
\affiliation{Quantum Information Science Group, Computational Sciences and Engineering Division, Oak Ridge National Laboratory, Oak Ridge, Tennessee 37831, USA}
\affiliation{Bredesen Center for Interdisciplinary Research and Graduate Education, University of Tennessee, Knoxville, Tennessee 37996, USA}


\begin{abstract}
This manuscript investigates the potential effect of a nuclear-disturbed atmospheric environment on the signal attenuation of a ground/satellite transmitter/receiver system for both classical optical and quantum communications applications. Attenuation of a signal transmitted through the rising nuclear cloud and the subsequently transported debris is modeled climatologically for surface-level detonations of 10~kt, 100~kt, and 1~Mt. Attenuation statistics were collected as a function of time after detonation. These loss terms were compared to normal loss sources such as clouds, smoke from fires, and clear sky operation. Finally, the loss was related to the degradation of transmitted entanglement derived from Bayesian Mean estimation. 
\end{abstract}

\maketitle



\section{Introduction}
\label{sec:level1}

The detonation of a nuclear weapon in the atmosphere can create adverse conditions for free space optical communications, including quantum communications. A nuclear detonation has three main outputs: blast, thermal and ionizing radiation~\cite{glasstone1977effects}. These effects occur over the span of seconds, but the resulting environment, i.e., the nuclear disturbed environment, can persist for hours to days afterwards. In a nuclear disturbed environment, two major effects can degrade an optical communications link: increased transmission loss of the optical signal and the addition of more background noise in the detector. The degradation of the optical link in a nuclear disturbed environment is caused by absorption/scattering of the optical signal due to nuclear cloud formation, weapon debris/soil lofting and elevated levels of particulate matter in the air (haze/smoke), for example, from a nearby burning city. Ionospheric effects from high altitude nuclear detonations can have significant effects (polarization, reflection, decoherence, phase effects, etc.~\cite{glasstone1977effects}) on a radio-frequency communications link due to the increased electron density in the upper atmosphere, but it has negligible effects in the near-IR optical range. In this paper, we consider near-IR wavelengths for the classical and quantum communications links. In this regime, the dominant loss mechanism will occur from absorption/scattering, and more specifically, Mie scattering due to the various aerosols and particulate matter injected into the atmosphere from the nuclear detonation. To keep the scope manageable, we consider detector noise effects elsewhere. 

As a nuclear fireball rises and cools, it forms a cloud filled with water vapor droplets, weapon debris and entrained material (soil, concrete, etc.) that was in the vicinity of the fireball. This cloud will cause significant scattering and absorption to the wavelengths we are investigating for the optical link. Additionally, the material lofted by the cloud will remain airborne for hours to days afterwards, depending on the particle size distribution of the material. The high thermal output of a weapon detonation can cause secondary fires to start in an urban or rural area. The subject of burning cities and `nuclear winter' has been studied in detail by many research groups over the past couple of decades~\cite{turco1990climate, penner1986uncertainties} but is outside the research scope of this study. Instead, we investigate blast-generated particulate matter in the air and its resulting effect on an optical communications link. The radiative output of the debris is substantial in the first hours, but the density of the lower atmosphere (generally around the tropopause and below for surface-level detonations) allows for charge recombination, minimizing the effect of ionization on beam transmission. We therefore assume this term is minor relative to the attenuation of the debris itself and neglect it.  In this way, we provide a scoping study of the additional loss that arises from various nuclear blast scenarios on an earth to satellite optical (conventional and quantum) communications link. 

Typical free-space conventional communications build in generous power margins by launching bright signals with millions or more of photons to account for signal losses. However, quantum communications generally operate at the single-photon level. Thus, while we compute the impact of loss generally, we expect the nuclear disturbed environment to present greater challenges for quantum communications. However, there are compelling capabilities not available without quantum signal transmission, namely information theoretic security through quantum key distribution (QKD) protocols~\cite{bennett2020quantum, Ekert1991QuantumCB, PhysRevLett.92.057901, 8527822}, where security remains independent of computational capability. In addition, by distributing entanglement, one can teleport quantum information over large distances provided one also has classical communications~\cite{bennett1993teleporting}. As it is required for teleportation, entanglement is considered a key resource for quantum communications and networking. Limitations on the range over which entanglement can be shared due to channel loss restrict the scale of global quantum networks~\cite{lucamarini2018overcoming}. Without quantum repeaters~\cite{PhysRevLett.81.5932}, a channel using commercial optical fiber can transmit quantum signals roughly only a couple of hundred kilometers~\cite{PhysRevLett.124.070501}. A terrestrial free-space optical (FSO) channel can suffer from obstacles in the field of view, atmosphere related effects, and the curvature of the earth that limit transmission range~\cite{malik2015free}. Satellite-based networks are a potential solution for long-range entanglement distribution on a global scale. Satellite-based architecture offers longer communication range than optical fiber due to lower attenuation scaling. Proof-of-concept demonstrations have been performed or are in conceptual phases for satellite quantum communications in China~\cite{liao2017satellite, ren2017ground}, the United States~\cite{hamiltona2019overview, seas2019optical}, the European Union~\cite{vallone2015experimental, ursin2009space}, Japan~\cite{takenaka2017satellite, Yamamoto_2019}, and Canada~\cite{Sussman_2019, jennewein2014qeyssat}. In addition, there has been much theoretical work analyzing the effect of the atmosphere and turbulence on satellite quantum communications~\cite{Aspelmeyer, Bonato_2009, doi:10.1177/1548512916684562, sharma2019analysis, Liorni_2019}. A feasibility analysis of satellite constellations to realize a global quantum network has been performed~\cite{VERGOOSSEN2020164}.

The main contribution of this manuscript is to estimate the additive nuclear disturbed penalty for earth-space communications links. That is, we calculate the loss penalty to an optical signal attempting to pass through (a) the stabilized nuclear cloud, (b) later-time transported airborne debris, and (c) ash and smoke from secondary fires initiated by the detonation. We assume that these loss terms are fully separable from the normal transmission losses in the atmosphere due to air, ambient airborne particulates, turbulence, etc., and that these loss terms (in dB) may be added to the atmospheric losses for the full transmitter-to-receiver attenuation estimate. We also assume that any photon interaction with the nuclear cloud or debris results in extinction of the photon (that is, no photon interacts weakly yet still survives to the receiver). Therefore, the loss terms are direction-independent and apply equally to both ground-to-satellite and satellite-to-ground transmission. All direction- and distance-dependent losses would be attributable to the atmospheric attenuation term.

\section{Background}

\begin{figure}[ht]
	\centering\includegraphics[width=.65\textwidth]{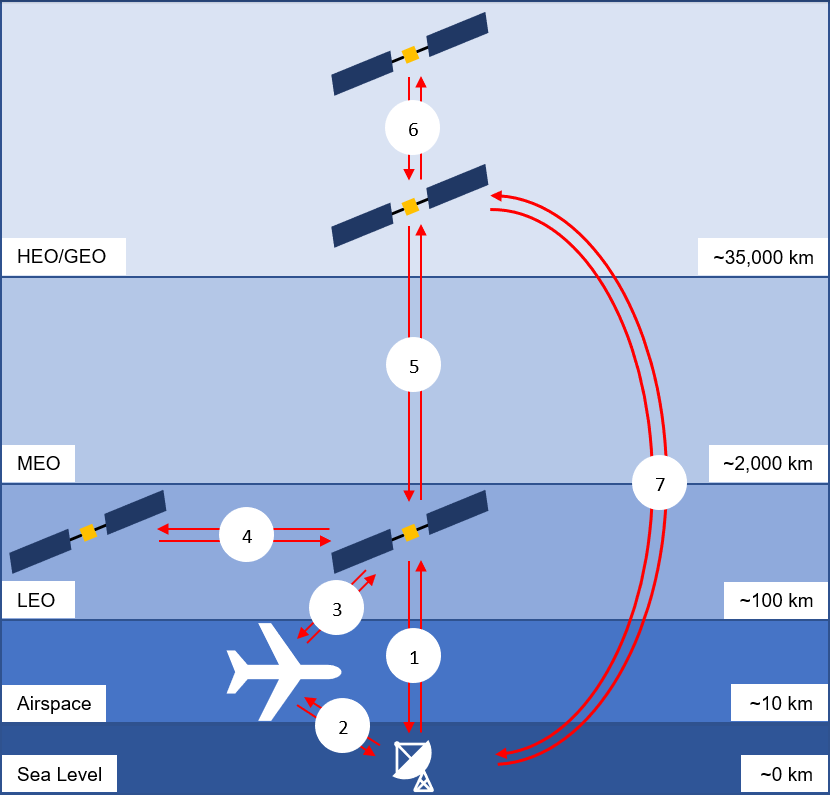}
	\caption{A schematic of the principle of operation for different link scenarios in a hypothetical space-based quantum and/or classical communications network: 1) Ground-to-low Earth orbit (LEO) 2) Ground-to-aircraft 3) LEO-to-aircraft 4) LEO-to-LEO 5) LEO-to-geosynchronous orbit (GEO) 6) GEO-to-GEO 7) Ground-to-GEO. All types of links can be bidirectional.} 
	\label{fig:PoO_Schematic}
\end{figure}

Before analyzing the performance of a quantum channel in a disturbed environment, it is necessary to determine what is possible during more favorable operating conditions. A concept of operations schematic for a future space-based communications network is shown in Fig.~\ref{fig:PoO_Schematic}. The figure provides a summary of the optical channels that can be considered and for concreteness, we specifically consider heralded two-qubit entanglement distribution between ground stations and satellites. In heralding, one photon of the entangled pair is measured at the transmitter while the other is sent over the free-space optical channel. In the ideal limit, this is analogous to having an ideal deterministic single photon source, and could also represent a directly encoded photon, as typically used in QKD.

An encoded quantum photon propagating through a free space optical (FSO) channel displays similar behavior as a classical pulse of light with otherwise similar characteristics (e.g., wavelength, spatial mode, pulse duration, etc.) transmitted in the same conditions. Modeling the optical loss of a laser-based FSO link therefore has become a standard way to gain insight into the performance  of a quantum link. More specifically, Gaussian beam propagation is applied to analyze the performance of an optical channel to infer the limitations of a quantum link due to diffraction and the atmospheric turbulence~\cite{Aspelmeyer}. This section introduces the Gaussian beam propagation methodology and the atmospheric model that analyze the optical channel of a quantum link between ground stations and satellites.

\subsection{Atmospheric Model} \label{Atm_Model}

A Gaussian beam is a focused or collimated beam of light where the irradiance and transverse electromagnetic field are defined by a Gaussian function. The Gaussian beam approximation is suitable for small divergence angles where the paraxial approximation is applicable. We assume that our laser beam is perfectly collimated and of a Gaussian form described by the fundamental transverse electromagnetic mode (TEM$_{00}$). The complex amplitude of the electric field for a Gaussian beam at a distance \textit{z} from the waist is given in~\cite{doi:https://doi.org/10.1002/0471213748.ch3} as
\begin{equation}
    U(\textbf{r}) = U_{0} \frac{\omega_{0}}{\omega(z)} \exp{\left(-\frac{r^{2}}{\omega^{2}(z)} \right)}\exp{\left(-ikz - ik\frac{r^{2}}{2R(z)} + i\zeta(z) \right)},
\end{equation}
where $U_{0}$ is the maximum value of the amplitude, $\omega_{0}$ is minimum beam radius at the waist (\textit{z} = 0), \textit{R}(z) is the radius of curvature, \textit{r} is the radius at a distance z, \textit{i} is the imaginary unit, \textit{k} is the wavenumber $k = 2\pi / \lambda$. The evolution of the real part of the Gaussian beam in the propagating direction $z$ can be calculated from the beam waist~\cite{alda2003laser}
\begin{equation}
    \omega(z) = \sqrt{\omega^{2}_{0} + z^{2}\theta^{2}_{d}}\ ,
    \label{eq:waist}
\end{equation}
where the beam divergence $\theta_{d}$ is given by
\begin{equation}
    \theta_{d} = \frac{\lambda}{\pi \omega_{0}}\ ,
    \label{eq:div}   
\end{equation}
and $\lambda$ is the channel wavelength. We assume the beam is transmitted either from a satellite or a ground station and the spot size spreads as it propagates to the receiver at another satellite or ground station. 

The irradiance of the beam is obtained from the modulus squared of the electric field as
\begin{equation}
    I(r,z) = I_{0}\frac{\omega^{2}_{0}}{\omega^{2}(z)} \exp{\left(-\frac{2r^{2}}{\omega^{2}(z)} \right)}\ .
\end{equation}
By integrating the irradiance an expression for the power passing through a circular receiver of radius $r_{0}$ is
\begin{equation}
   P(z) = \int_{0}^{2\pi} \int_0^{r_{0}} I(r,z)r \mathrm{d}r \mathrm{d}\theta\ .
   \label{eq:Power_int}
\end{equation}
By solving Equation \ref{eq:Power_int}, the transmission power ratio for an unfocused beam is
\begin{equation}
    \frac{P_R}{P_T} = 1 - \exp \left(- \frac{2r^{2}}{\omega^{2}(z)} \right).
    \label{eq:trans}
\end{equation}
The link efficiency ratio for an channel is the power passing through a receiver aperture of diameter $D_R$ from a transmitting device of diameter $D_T$ that is separated by a link distance of $L$ and is calculated as
\begin{equation}
    \frac{P_R}{P_T} = 1 - \exp\left(-\frac{2D^2_R}{D^2_T + 4L^2 \theta_{d}}\right).
    \label{eq:Frac_Power}
\end{equation}
Here, the minimum beam waist is treated as half of the diameter of the transmitting aperture $\omega_{0} = D_{T}/2$ and the collection radius is half of the diameter of the receiving aperture $r = D_{R}/2$. 
The optical link attenuation is the inverse of the link efficiency:
\begin{equation}
    \frac{P_T}{P_R} = \left(1 - \exp \left(-\frac{2D^2_R}{D^2_T + 4L^2 \theta^{2}_{d}}\right)\right)^{-1}.
    \label{eq:atten_unitless}
\end{equation}
Equation \ref{eq:atten_unitless} is sufficient to understand the propagation of a beam that is purely diffraction limited and gives a lower bound of the channel attenuation. Given the typically high loss over the scenarios considered, the attenuation \textit{A} is more conveniently expressed in dB by converting the attenuation ratio as follows:
\begin{equation}
    A = 10 \times \log_{10} \left(\frac{P_T}{P_R}\right).
\end{equation}

Several obstacles complicate any attempt to transmit an optical signal through the atmosphere including scattering, absorption, and variations in the atmospheric refractive index~\cite{doi:10.1177/1548512916684562}. The atmosphere consists of various gases, atoms, water vapor, and molecules trapped by the gravity of the earth. The various molecules present in the atmosphere can absorb or scatter a photon leading to a reduction in signal power. Meanwhile, changes to the temperature, density, pressure, and humidity of the air cause variations in the refractive index of the atmosphere~\cite{andrews2005laser}. The propagation of a beam through the atmosphere causes a distortion of the beam shape due to the constant variation of the index of refraction. These effects manifest as twinkling, beam wandering, image distortion, etc.~\cite{fante1975}. Various mathematical models have been proposed that attempt to characterize the index of refraction structure constant~\cite{kaushal2015free}.

We model the index of refraction structure constant, $C^{2}_{n}$, as a function of altitude, $h$, using the Hufnagel-Valley (HV) model~\cite{andrews2005laser}. The HV model was proposed by Hufnagel based on balloon-borne measurements of the temperature of the troposphere made by Bufton et al.~\cite{hufnagel1974variation, hufnagel1978propagation, bufton1972measurements}. The model was improved later by Valley with the inclusion of a ground-level turbulence term~\cite{valley1980isoplanatic} that increases the scope of the different conditions on the surface~\cite{sasiela2012electromagnetic}. The HV model is chosen due to the flexibility offered by the parameters to model the turbulence in different geographical locations. The atmospheric structure constant according to the HV model is:
\begin{equation}
\begin{aligned}
    C^2_n (h) = \, \, 5.94 \times 10^{-53} \left( \frac{v}{27} \right)^2 h^{10} e^{-\frac{h}{1000}} + 2.7 \times 10^{-16} e^{-\frac{h}{1500}} + C^2_n(0)e^{-\frac{h}{100}}\ .
    \label{eq:HV}
\end{aligned}
\end{equation}
Here, the atmospheric structure constant is a function of the following: increasing height from the surface, \textit{h}, the root mean squared wind speed, \textit{v}, and the ground level turbulence, $C^2_n(0)$. We chose the typical values of \textit{v}~=~21 m/s for the wind speed velocity and a ground-level turbulence of $C^2_n(0) = 1.7 \times 10^{-14}$ m$^{-2/3}$ to obtain an approximation of the atmospheric turbulence where the isoplanatic angle is 7 $\mu$rad and the atmospheric coherent length is 0.05 m at $\lambda$ =~500 nm.

Using this form of the structure constant, the effect of the atmospheric turbulence on a propagating beam is determined using the atmospheric coherence length or the Fried parameter, $r_{0}$. The Fried parameter is defined in~\cite{Fried:66} as
\begin{equation}
    r_0 = \left(0.423k^2 \sec{\zeta} \int^L_{h_0} C^2_n(z') \mathrm{d}z' \right)^{-\frac{3}{5}}.
    \label{eq:Fried}
\end{equation}
where $\zeta$ is the zenith angle, \textit{k} is the wave number $k = 2\pi/\lambda$, and $h_{0}$ is the starting height. The divergence angle as a result of atmospheric turbulence is
\begin{equation}
    \theta_{atm} = \frac{\lambda}{r_{0}}\ .
    \label{eq:atm_divg}
\end{equation}
We assume that the divergence due to the atmosphere is squared and additive to the diffraction divergence~\cite{Aspelmeyer} due to the divergences being dependent on different sources. The attenuation for an optical link through the atmosphere is given in~\cite{Aspelmeyer} as
\begin{equation}
    \frac{P_T}{P_R} = 10^{\frac{A_{air}}{10}}\left(1 - \exp \left(-\frac{2D^2_R}{D^2_T + 4L^2 (\theta^{2}_{d} + \theta^{2}_{atm})}\right)\right)^{-1},
    \label{eq:atten_atm}
\end{equation}
where $A_{air}$ (dB) is the attenuation of the atmosphere due to scattering and absorption. (Note that~\cite{Aspelmeyer} expressed air attenuation as $A_{atm}$; we changed this to $A_{air}$ and instead use $A_{atm}$ to refer to attenuation due to air molecules, turbulence, and thermal diffraction within the atmosphere.)

We are only interested in the fundamentals of the problem and therefore, we do not consider pointing loss, transmitting factors for the telescopes, or detector inefficiency. These are scaling factors and applications of these parameters to the equations can be found here~\cite{Aspelmeyer,sharma2019analysis}. Calculations using Equations \ref{eq:atten_unitless} and \ref{eq:atten_atm} were verified by comparing to previously published results of Aspelmeyer et. al~\cite{Aspelmeyer}.

The optical channel is assumed to operate at a wavelength of 1550~nm, which has an atmospheric attenuation $A_{air} = $ 1~dB from~\cite{Aspelmeyer} for scattering and absorption by air molecules between ground and space. The 1550~nm wavelength is assumed due to existing satellite technology operating in that wavelength range and eye safety concerns of using shorter wavelengths. A link distance of $L = $ 400~km is assumed between the ground station (at sea-level, $h_{0} = $ 0~km) and orbiting satellite. The space aperture has an assumed diameter of 0.1~m ($D_T$ for downlink, and $D_R$ for uplink) and the ground station aperture has an assumed diameter of 1~m ($D_R$ for downlink, and $D_T$ for uplink). The angular divergence is $\theta_{d} = 9.87 \times 10^{-7}$ for the uplink scenario and $\theta_{d} = 9.87 \times 10^{-6}$ for the downlink case. In the uplink case, the Fried parameter is $r_{0} = $ 0.193~m and the atmospheric divergence is $\theta_{atm} = 8.04 \times 10^{-5}$. Using Eq.~\ref{eq:atten_atm}, the downlink and uplink clear air attenuation for this configuration are 16.0~dB and 34.0~dB, respectively. This gives us a reference optical link scenario, which we will later consider in the context of how a heralded entangled link degrades with the addition of the nuclear disturbed attenuation.

For debris particles, all scattering and absorption interactions are assumed to remove a photon from the optical beam, so the extinction attenuation (the sum of the scattering and absorption attenuations) $Q_{ext} = Q_{a} + Q_{s}$ is taken as the metric for beam loss due to debris interference. Because the larger particle sizes fall out of the atmosphere relatively quickly, Mie scattering is assumed for all airborne particles at 1~hour or later. The PyMieScatt Python package~\cite{PyMieScatt-2018} is used to calculate $Q_{ext}$~\cite{Mie-1908} between the transmitter and receiver due to all particle sizes at all atmospheric levels using the equation

\begin{equation}
    Q_{ext} = \frac{2}{x^2}\displaystyle\sum_{n=1}^{n_{max}}(2n+1)\operatorname{Re}(a_n+b_n)\ ,
    \label{Eq:Q_ext}
\end{equation}
where $a_n$ and $b_n$ are coefficients that are functions of the complex refractive index $m$ and the size parameter 
\begin{equation}
    x={\pi}d_p/\lambda\ ,
    \label{Eq:size_parameter}
\end{equation}
where $d_p$ is the particle diameter and $\lambda$ is the wavelength. Note that this does not yet include attenuation from the atmosphere. Attenuations are calculated for 800~nm and 1550~nm wavelengths. By evaluating Eq.~\ref{Eq:Q_ext} for all particle sizes, the \mbox{per-particle} extinction attenuation is found for all particles.

The total attenuation (in dB) of an optical transmission between a ground station and a satellite is calculated as the integral of the attenuation of all particles between the transmitting and receiving stations. The discretization of the atmosphere within the transport calculations renders the integral as a sum over the vertical layers. The total attenuation is therefore given as
\begin{equation}
    A_{total} = \displaystyle\sum_{l=1}^{l_{max}}\displaystyle\sum_{d=1}^{100} \left[ 10* \log_{10}(e^{n_{d,l}*\pi*r_d^2*Q_{ext,d}*V_l}) \right]\ ,
    \label{Eq:Atten_total}
\end{equation}
where $l$ is the vertical layer, $d$ is the particle size class, $n_{d,l}$ is the vertical particle number density (particles per meter), $r_d$ is the particle radius, $Q_{ext,d}$ is the single-particle extinction attenuation, and $V_l$ is the layer thickness.

For this simulation, the attenuation is independent of the direction of the signal transmission (i.e., ground-to-satellite versus satellite-to-ground). This independence is the result of two simplifying assumptions. First, \emph{only} non-interacting photons pass through the debris. Any interaction at all results in photon extinction. This renders the problem tractable, as solutions do not need to be considered for the rare cases of photons that may weakly interact yet still reach the receiver with some altered angle and wavelength. Second (and partly a result of the first assumption), the signal attenuation is separable between the debris and the atmosphere. The attenuation effects of atmospheric diffusion ($A_{atm}$), signal direction ($A_{dir}$), and aperture sizes ($A_{aperture}$) are also assumed to be additive to the attenuation of the cloud ($A_{cloud}$), debris ($A_{nuc}$), or smoke ($A_{smoke}$) in the attenuation calculations that follow using Equation~\ref{Eq:dB_add}. This should hold true because the signal is weak and operating on a single photon level so that nonlinear effects do not need to be considered.

\begin{equation}
    A_{total} = 10\log_{10}{(10^{\frac{A_{atm}}{10}} + 10^{\frac{A_{nuc}}{10}} + 10^{\frac{A_{dir}}{10}} + \cdots{})}\ .
    \label{Eq:dB_add}
\end{equation}

Further, the attenuation dependence on transmitter/receiver distance as well as the direction of transmission (i.e., satellite-to-ground or ground-to-satellite) are contained entirely within $A_{atm}$, $A_{dir}$, and $A_{aperture}$~\cite{Aspelmeyer}. Since the nuclear disturbance terms $A_{nuc}$, $A_{cloud}$, and $A_{smoke}$ are assumed to be extinction-only attenuation effects, their resulting losses are the same whether the material is close to the transmitter, close to the receiver, or anywhere in between. Therefore, the nuclear-disturbed attenuation losses apply equally to ground-to-satellite communications and satellite-to-ground communications, regardless of the altitude of the satellite. These terms also apply equally to ground-to-aircraft and aircraft-to-ground communications if the debris is assumed to be entirely within the line of sight (i.e., the aircraft is flying above the debris, not through or below it).

\subsection{Quantum State Estimation (QSE)}\label{QSE}
To understand how a quantum link will degrade, we must consider how the reduced transmission impacts our ability to estimate the quality of a transmitted quantum state.  Our quantum state estimation technique utilizes statistical inference via application of Bayes rule,
\begin{equation}
    P(\tau | \mathcal{D}) = \frac{P(\mathcal{D}|\tau)P(\tau)}{P(\mathcal{D})},
    \label{Eq:Bayes_Rule}    
\end{equation} 
 which allows expression of the probability of parameter set $\tau$ given dataset $\mathcal{D}$, $P(\tau | \mathcal{D})$, based on specific events and prior knowledge. In this case, our events are photon counts measured from a photon detector. Our parameter space describes the quantum state as well as experimental aspects such as photon attenuation. We previously applied this technique to two-qubit states in some simple scenarios~\cite{williams2017quantum}.

Our specific concept of operations involves an entangled photon pair source operated as a heralded single-photon source, a single loss transmission channel, and a perfect measurement apparatus. The primary parameter affecting the state estimation is the attenuation of the channel.

\begin{figure}[htp]
	\centering\includegraphics[width=.65\textwidth]{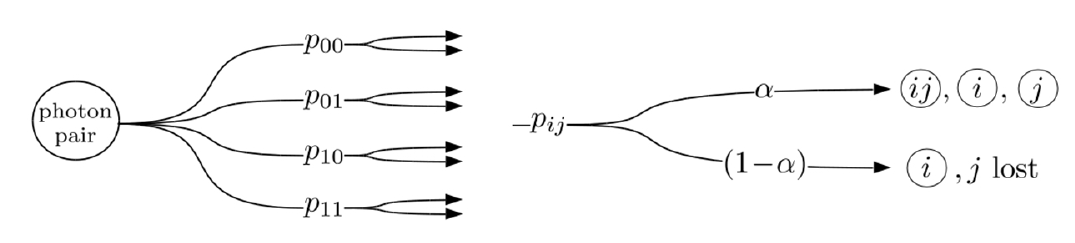}
	\caption{Potential measurement and loss outcomes for sender and receiver's shared photon pair.} 
    \label{fig:Tree}
\end{figure}

The application of Bayes's rule to a single state channel includes complete enumeration of the possible experimental outcomes given our initial state, a single pair of entangled photons. For the heralded single-photon source, we assume perfect detection of the sender's herald photon, imperfect transmission of the signal photon sent to the receiver, and perfect measurement and detection by receiver's apparatus. Fig.~\ref{fig:Tree} illustrates the potential outcomes for our photonic qubits in any given measurement basis in which both sender and receiver measure a binary result, 0 or 1, with a transmission channel efficiency $\alpha$, $1-\alpha$ is then the attenuation.

We can use this probability tree to build the single-basis likelihood. In this scenario, the single-basis likelihood is

\begin{equation}
\begin{aligned}
    P_{kl}( \mathcal{D}| p_{ij}, \alpha) =\,\, \alpha^{C}(1-\alpha)^{A_0 + A_1 - C} 
    \times p_{00}^{c_{00}} p_{01}^{c_{01}} p_{10}^{c_{10}} p_{11}^{c_{11}} 
    \times (p_{00} + p_{01})^{A_{0}-c_{00}-c_{01}} 
    \times (p_{10} + p_{11})^{A_{1}-c_{10}-c_{11}}\ ,
    \label{eq:prob_tree}
\end{aligned}
\end{equation}

where $k, l \in {I,X,Y}$, to denote measurements in the rectilinear, diagonal, and circular bases, respectively, $\mathcal{D} = \{A_{0}, A_{1}, c_{00}, c_{01}, c_{10}, c_{11}\}$, $\mathcal{C} = c_{00} + c_{01} + c_{10} + c_{11}$, and $A_{0} + A_{1} = N$ being the total number of photon pairs shared between sender and receiver. $A_{0}$ and $A_{1}$ are sender's single-photon counts and $c_{00}$, $c_{01}$, $c_{10}$, and $c_{11}$ are the sender's and receiver's coincidence counts. As can be seen in Eq. \ref{eq:prob_tree}, the attenuation is the same for all bases and measurement channels which leads to the reduced representation in which the attenuation is not intertwined with the individual measurement outcomes. Thus, the attenuation's effect on the quantum state estimation, in this case, amounts to an increased uncertainty due to fewer measurements results.  Understanding this uncertainty is important as one is often required to do parameter estimation to carry out a desired protocol, for example, to do quantum key distribution.  The complete likelihood can then be written as a product of the individual basis likelihoods,

\begin{equation}
    P(\tau,\alpha| \mathcal{D}) = \displaystyle\prod_{k,l} P_{k,l}(\tau,\alpha | \mathcal{D})\ ,
\end{equation}
where we have adjusted the parameterization ($p_{ij}\rightarrow\tau$), see Appendix \ref{densityMatrixAppendix}, such that the density matrix ($\rho$) associated with these probabilities fulfills all quantum criteria:

\begin{center}
\begin{tabular}{rl}
    $Tr(\rho) = 1$ &Probabilities sum to 1,    \\
    $\left\langle \psi \middle| \rho \middle| \psi \right\rangle \geq 0$ &Positive semi-definite, \\
    $\rho = \rho^{\dagger}$ &Hermitian.
\end{tabular}
\end{center}

With the complete parameterized likelihood, we can apply Bayes's rule to obtain the normalized probability distribution. With this we simply integrate to estimate any mean or related standard deviation of any parameter or function of parameters $\mathcal{F}(\tau,\alpha)$ as
\begin{equation}
    \bar{\mathcal{F}} = \int\int \mathrm{d}\tau\mathrm{d}\alpha P(\tau,\alpha | \mathcal{D}) \mathcal{F}(\tau,\alpha)\label{BMEintegral}\ .
\end{equation}
The complexity of the integral typically requires a numerical approach to evaluate it. We use a Markov chain Monte Carlo algorithm known as slice sampling~\cite{neal2003slice} to evaluate the integrals. Thus, calculation of any integral of the form given in Eq. \ref{BMEintegral} is approximated numerically as
\begin{equation}
    \bar{\mathcal{F}}\;= \lim_{R\rightarrow\infty}\frac{1}{R}\sum_{i=1}^{R} \mathcal{F}^{(r)}\ ,
    \label{PDlast}
\end{equation}
where $\mathcal{F}^{(r)}$ is a single value $\mathcal{F}(\tau^{(r)},\alpha^{(r)})$ at specific sample points $\tau^{(r)}$ and $\alpha^{(r)}$. This approximation converges on the correct value for large $R$. In our approach, independent samplers are run in parallel. The statistics from each sampler are then compared with each other. Once, all samplers agree, defined by their estimation overlap, the numerical approximation is accepted~\cite{williams2017quantum}. The quantity central to our entanglement analysis is the density matrix for which the mean estimate is
\begin{equation}
    \bar{\rho} = \int\int \mathrm{d}\tau\mathrm{d}\alpha P(\tau,\alpha | \mathcal{D}) \rho(\tau)\ \textrm{.}\label{BMEintegralRho}
\end{equation}
We note that the density matrix as defined in Appendix \ref{densityMatrixAppendix} is connected to the probability distribution $P(\tau,\alpha | \mathcal{D})$ through the parameters $\tau$.

There are a variety of entanglement measures and metrics used within the research community. Some of these have specific relevance to a certain application or perhaps an especially simple evaluation. We have chosen to use the Entanglement of Formation (EOF) as our entanglement measure which is based on another quantity called the \emph{concurrence} $\mathcal{C}$~\cite{wootters1998entanglement}. For two qubits, the concurrence is expressible as
\begin{equation}\mathcal{C}=\max\{0,\lambda_1-\lambda_2-\lambda_3-\lambda_4\}\ ,\end{equation}
where the $\lambda$'s are the decreasing $\left(\lambda_1\geq \lambda_2\geq \lambda_3\geq \lambda_4\right)$ eigenvalues of the matrix $\rho \cdot \tilde{\rho}$ with
\begin{equation}\tilde{\rho}=\left(\sigma_y \otimes \sigma_y\right)\rho^{*}\left(\sigma_y \otimes \sigma_y\right) \textrm{   with   } \sigma_y=
\textstyle{\begin{pmatrix}
0 & -i \\
i & 0
\end{pmatrix}}\ \textrm{.}\end{equation}
The EOF is then determined using the formula
\begin{equation}E\left(\rho\right)=h\textstyle{\left(\frac{1+\sqrt{1-\mathcal{C\left(\rho\right)}^2}}{2}\right)}\label{EOF}\ ,\end{equation}
with
\begin{equation}h\left(x\right)=-x\log_2(x)-(1-x)\log_2(1-x)\ ,\end{equation}
where h(x) is known as the binary entropy.


\subsection{Nuclear Debris Model}

The Defense Land Fallout Interpretive Code (DELFIC) is used to calculate the nuclear cloud dimensions as a function of time. DELFIC models nuclear cloud rise and formation by solving eight differential equations related to the detonation and the surrounding atmosphere~\cite{norment-1979-1,norment-1979-2}. The cloud rise model begins when the fireball reaches pressure equilibrium with the atmosphere and is essentially a rising hot gas bubble within a cooler medium. At this time, the shock wave has separated from the fireball, leaving temperatures of the gas phase and condensed phase, mass of debris and entrained air, internal kinetic energy (of the vorticity of the toroidal core), and the fireball rise rate, radius, and volume. Ballistic ejecta, which are typically very large particles (perhaps $>$1~cm in diameter), fall to the ground within a few seconds and are ignored. The remaining lofted debris is modeled as the DELFIC default lognormal particle size distribution for U.S. continental soil, which ranges from about 3~microns to 450~microns in diameter~\cite{norment-1979-1}. The conservation equations are solved iteratively in time until the cloud stabilizes at its final altitude, at which point it has lost its internal vorticity. During this time, the advection of the cloud due to ambient winds is accounted for, and the gravitational settling of particles is computed, allowing some of the larger particles may fall to the ground prior to the end of cloud rise. These calculations result in a description of all lofted particle sizes at cloud stabilization time and whether they remain in the cloud volume, have landed on the ground, or are at some intermediate altitude.

Once DELFIC has calculated the particle locations at cloud stabilization, these data are passed to the National Oceanic and Atmospheric Administration's (NOAA) Hybrid Single-Particle Lagrangian Integrated Trajectory (HYSPLIT) code for atmospheric transport~\cite{HYSPLIT-2016}. HYSPLIT's particle representation (as opposed to a puff representation) of the debris is transported using atmospheric and wind data from the North American Mesoscale Forecast System (NAM) at 12~km resolution~\cite{nam-2021}. The HYSPLIT's ``concentration'' calculations are used to predict the volumetric concentrations of all debris particle sizes in the atmosphere as a function of time after detonation, altitude, and latitude/longitude.

The DELFIC cloud rise model and the HYSPLIT atmospheric transport code are integrated using the ORNL-built Airborne Planning Tool (APTool)~\cite{lefebvre-2018b, hooper-2018}. A surface-level nuclear detonation is simulated every 4 hours from January 1 through December 31, 2019, resulting in over 2,000 simulations per yield scenario. The detonation simulations are independent of each other (i.e., not cumulative over time). The debris location as a function of latitude, longitude, and altitude is calculated at one-hour intervals from 1 to 6~hours at a latitude/longitude resolution of 0.05~degrees (very roughly 5~km) and eight to thirteen vertical layers, depending on the yield. The simulation ensemble is performed for yields of 10~kt, 100~kt, and 1~Mt (1,000~kt). Concentrations are calculated using 10~minute averages to reduce instantaneous spikes or depressions in concentration. The resulting concentration outputs are then inspected using APTool post-processing utilities to find the peak attenuation due to debris between a ground transmitter and a vertically-overhead receiver. As discussed in Section~\ref{Atm_Model}, these results apply equally to satellite-to-ground transmission. Signal losses (in dB) are assumed to be additive to normal atmospheric losses, so a zero loss condition would be expressed as 0~dB.

\section{Results}

\subsection{Nuclear Link Loss - Stabilized Cloud}

\begin{figure}[htp]
	\centering\includegraphics[]{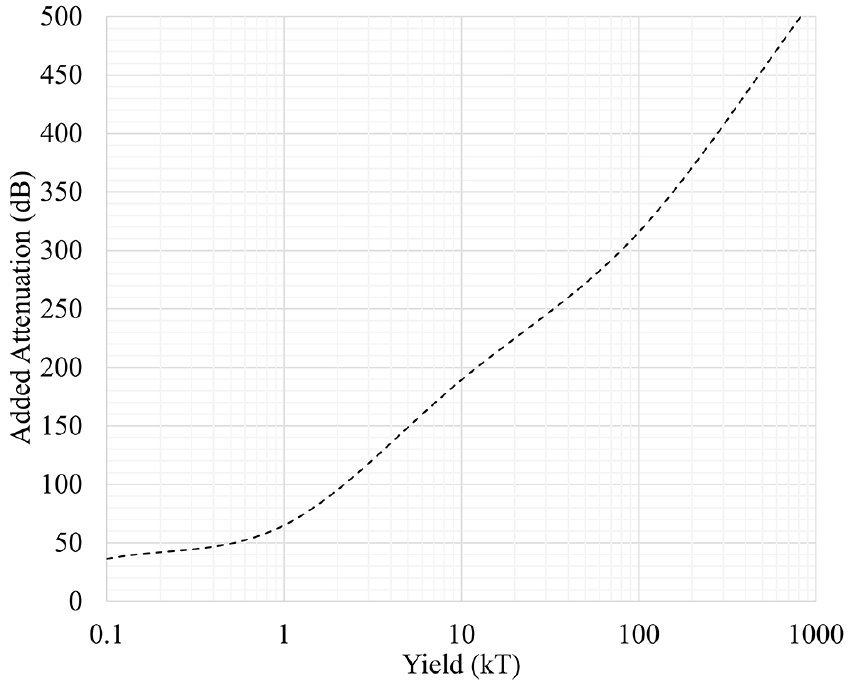}	
	\caption{1550 nm optical attenuation through a nuclear cloud as a function of weapon yield.} 
	\label{fig:Cloud_dB}
\end{figure}

For the first effect in a nuclear disturbed environment, nuclear cloud formation and stabilization are calculated for yields of 10~kt, 100~kt and 1~Mt at a height of burst of 1~m. The cloud was represented as a homogeneous mixture of water vapor (the liquid water content of the cloud is 1~g/m$^3$), soil and weapon debris. The particle size distribution for the soil and weapon debris is assumed to be the default DELFIC lognormal distribution with a median particle diameter around 130~$\mu$m~\cite{norment-1979-1} and the water vapor droplets are assumed to have a uniform diameter of 12~$\mu$m.  The cloud data is inputted into the MIE scattering equations~\cite{Mie-1908} to determine the amount of attenuation the optical link experienced during propagation through the nuclear cloud. The optical link is assumed to pass through the center of the nuclear cloud. The added attenuation through the stabilized cloud is given in Fig. \ref{fig:Cloud_dB}. While the actual stabilization time and size of the stabilized cloud will vary with atmospheric conditions for a given shot scenario (e.g., humidity, temperature, wind conditions, air pressure, etc.), very rough conceptual sizes and times for the stabilized clouds may be considered as follows: 3~km cloud diameter at 5~min for a 10~kt yield, 10~km cloud diameter at 10~min for 100~kt, and 30~km cloud diameter at 15~min for a 1~Mt yield. Even for a 1~Mt detonation, direct transmission through a stabilized cloud is unlikely, and even then only for a small window of time.

\subsection{Nuclear Link Loss - Transported Debris} \label{Debris_Atten}

During the initial cloud rise, the vorticity within the cloud causes it to retain shape, minimizing atmospheric diffusion. Additionally, the debris falling from the cloud is directly under the cloud itself and can result in very large attenuations, albeit for relatively small footprints. Eventually, as the vorticity weakens and the cloud cools, the cloud rise ceases and the debris begins to move in the atmosphere according to the winds, turbulent diffusion, and other meteorological effects. For a 1~Mt explosion, cloud rise typically ends about 15 to 20~minutes after detonation, while cloud rise from 100~kt and 10~kt detonations typically end after about 10~minutes and 5~minutes, respectively. From here, the debris will diffuse in the atmosphere, lowering the peak volumetric concentration and, by extension, the peak attenuation. Diffusion also causes the debris to occupy a larger volume of the atmosphere, increasing the region in which ground/satellite communications may be affected. Since the transport and diffusion is heavily dependent on atmospheric conditions, an ensemble of simulations is performed for each explosive yield value to generate a statistical measure of the potential signal attenuation as a function of time. 

\begin{figure}[htp]
\centering\subfloat[March 1, noon detonation]{%
  \centering\includegraphics[clip,width=0.8\columnwidth]{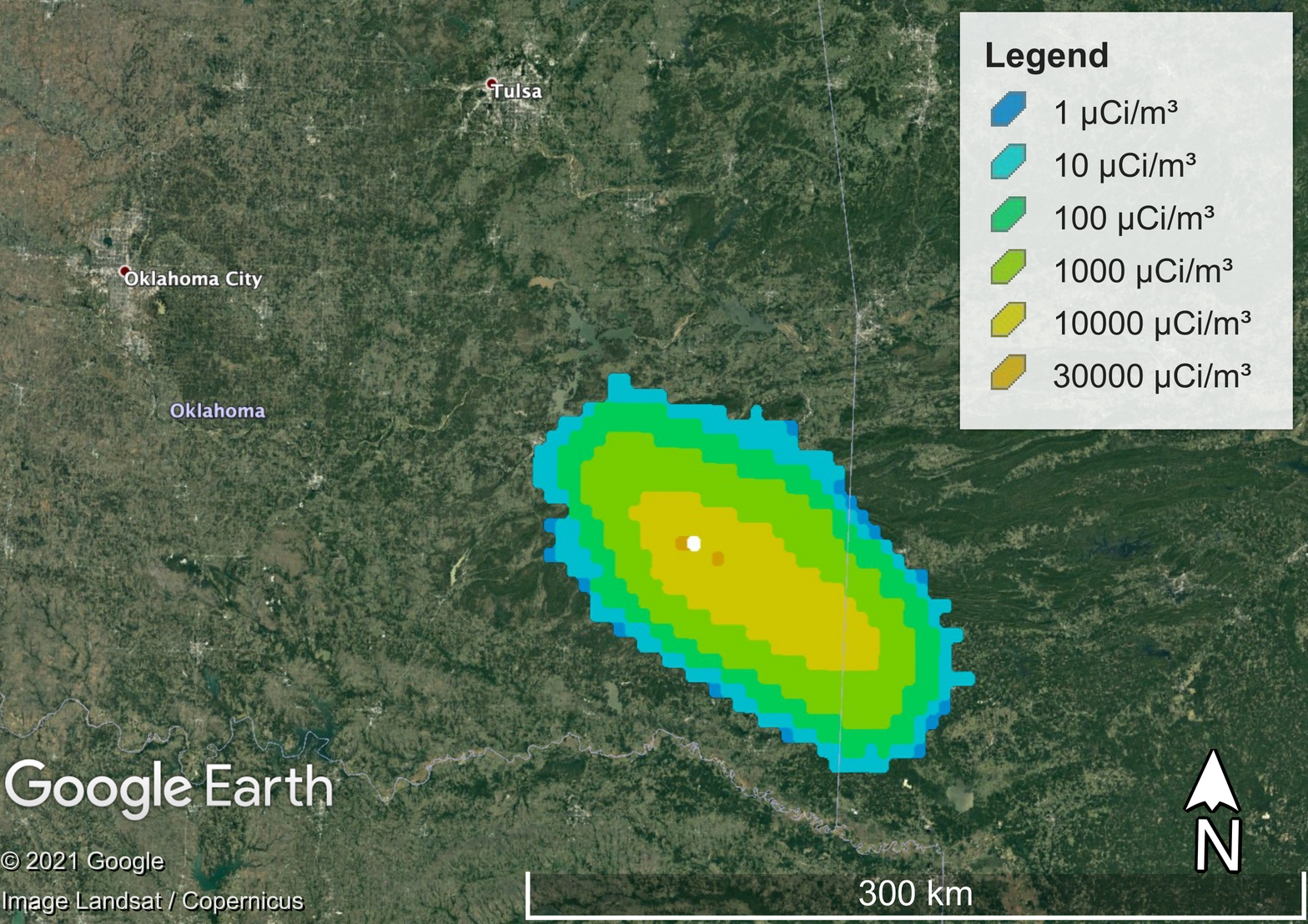}%
}

\centering\subfloat[October 1, 8 AM detonation]{%
  \centering\includegraphics[clip,width=0.8\columnwidth]{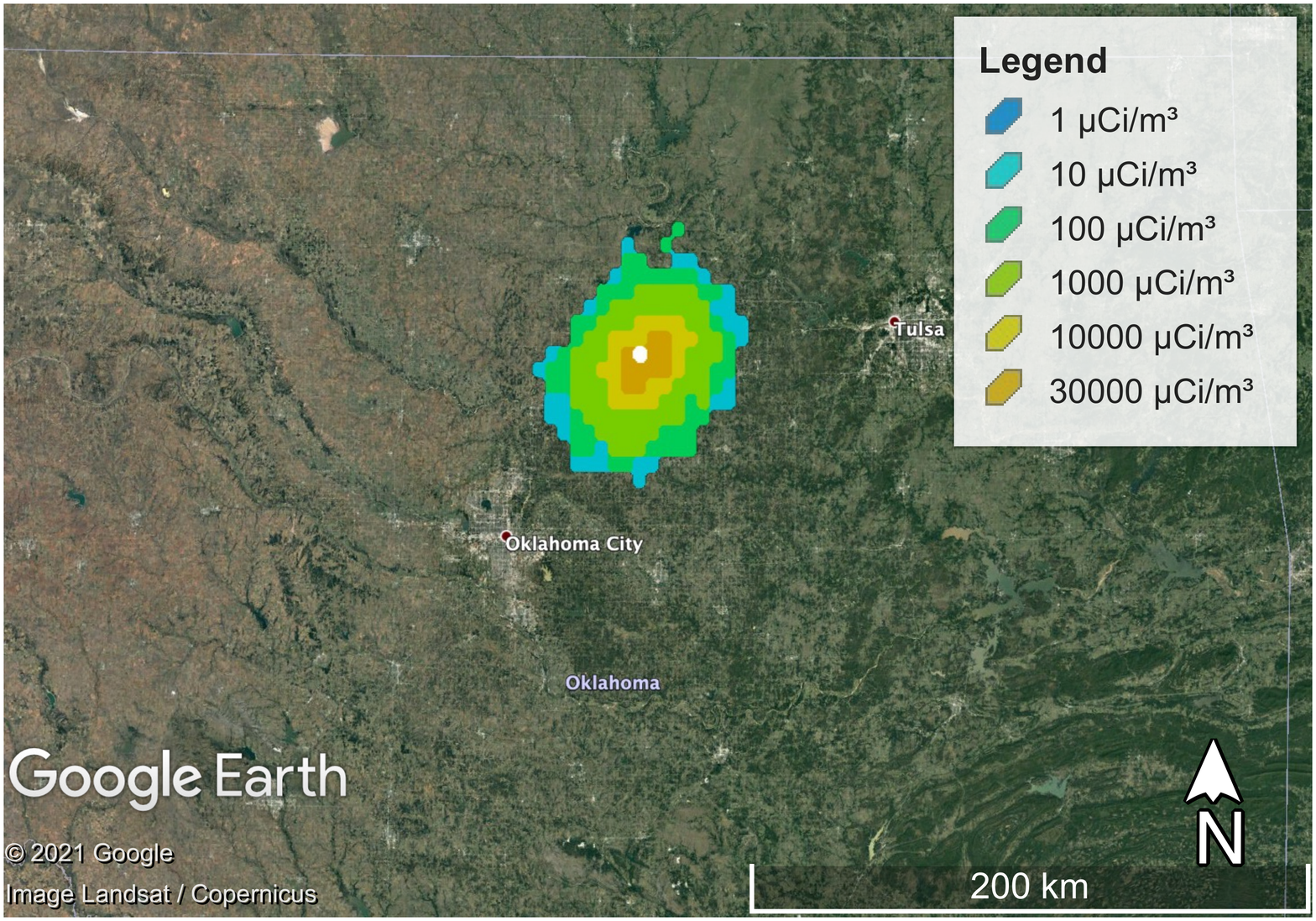}%
}

	\caption{Debris activity (in {\textmu}Ci/m\textsuperscript{3}) at 10~km above ground at 1~hr after a 1~Mt detonation in Oklahoma City for two different detonation times. The location of maximum attenuation is shown as a white dot.} 
	\label{fig: March_1hr}
\end{figure}

\begin{figure}[htp]
	\centering\includegraphics[width=.65\textwidth]{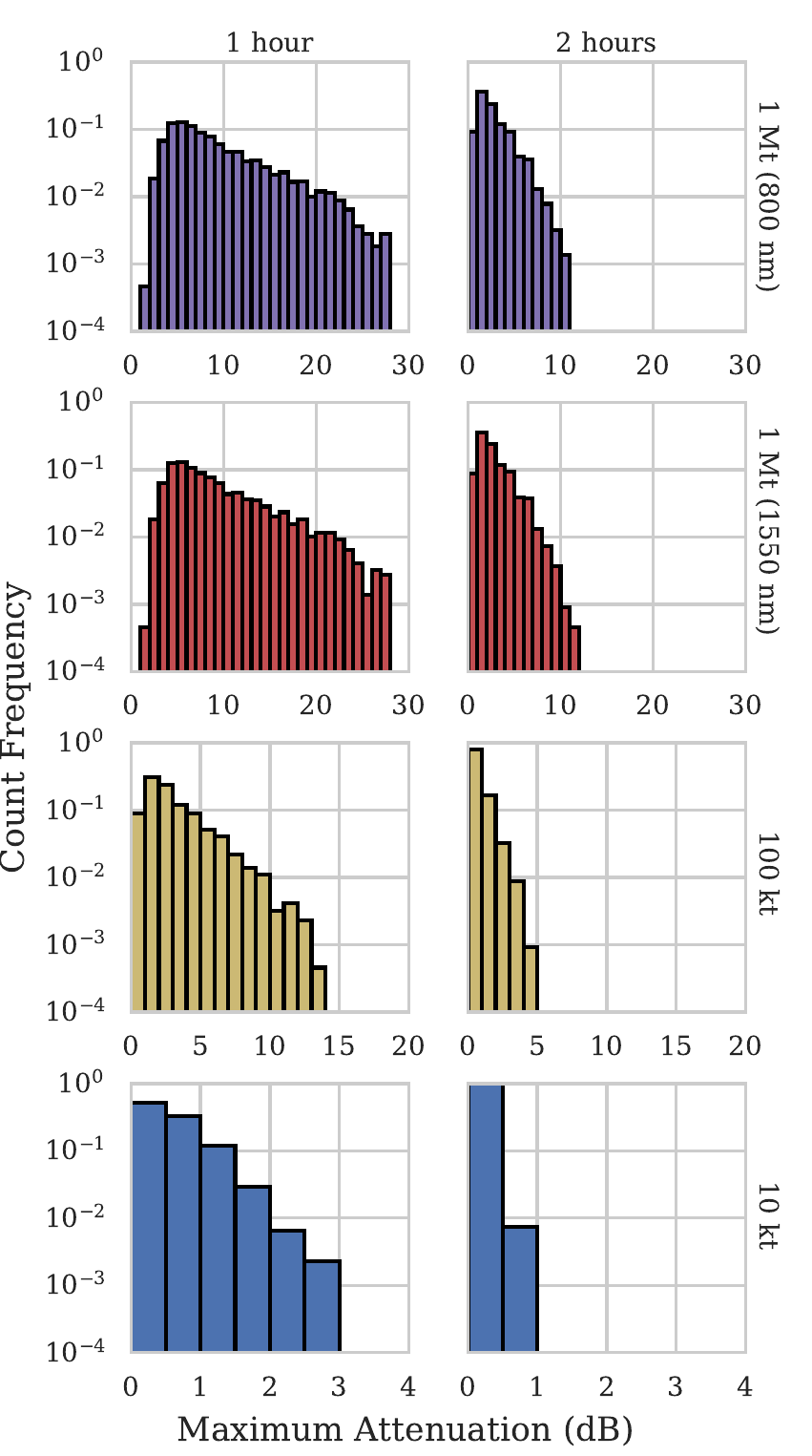}
	\caption{Maximum attenuation at 1 and 2~hours after detonation. 1~Mt attenuation is shown for both 800~nm and 1550~nm wavelengths for comparison. All others are 1550~nm.} 
	\label{fig: max_atten_all}
\end{figure}

The ensembles are analyzed in two manners. First, the maximum observed attenuation of each simulation is recorded as a function of time, regardless of downwind position. Fig.~\ref{fig: March_1hr} illustrates the change in location of the attenuation observation for two different detonation times. This represents the maximum possible signal loss that the nuclear debris could cause at a given time after detonation and is therefore an upper bound to potential signal disruption. The maxima are determined for all simulations within an ensemble and collected into a distribution of upper bounds as a function of time. Fig.~\ref{fig: max_atten_all} shows the aggregated maximum observed attenuation of each simulation for yields of 1~Mt, 100~kt, and 10~kt. The 1550~nm wavelength is evaluated for all yields, and the 800~nm wavelength is shown for the 1~Mt case as a comparison. For 10~kt yields, no attenuation greater than 3~dB is ever observed. For 100~kt, attenuation is only once observed as high as 14~dB at 1~hour, but 90\% of scenarios have a maximum attenuation of 5~dB or less, and 99\% have a maximum attenuation of 10~dB or less. At 1~Mt, 96\% of all scenarios have a maximum attenuation of 20~dB or less at 1~hour, and all but one scenario have a maximum attenuation of 10~dB by 2~hours. The 800~nm and 1550~nm simulations provide nearly identical results due to the wide range of particle sizes within the transported debris. Because of this, the rest of the results are given for the 1550~nm wavelength and are assumed to be applicable to the 800~nm wavelength.

\begin{figure}[htp]
	\centering\includegraphics[width=.55\textwidth]{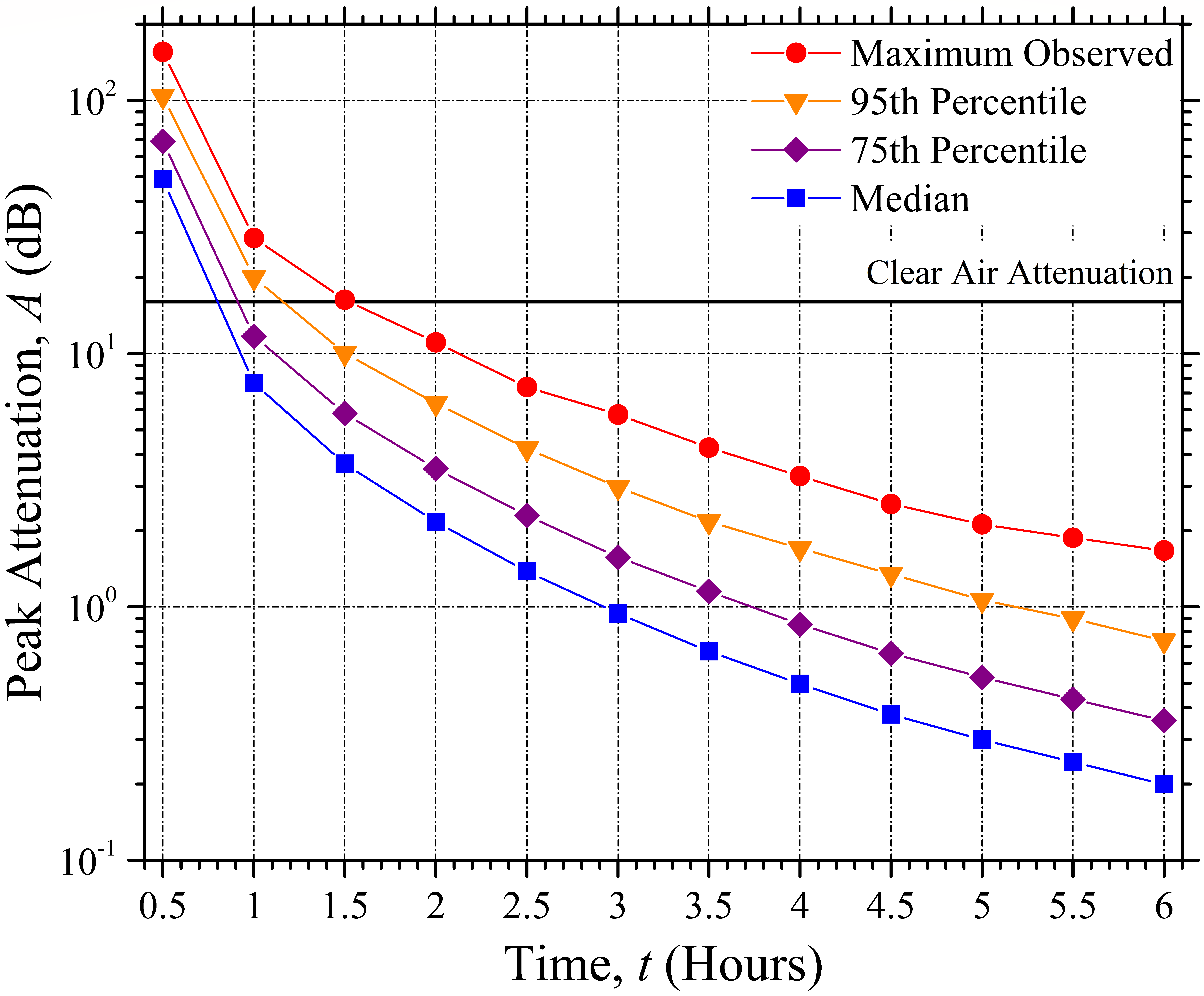}
	\caption{Maximum 1550~nm attenuation at 1~Mt.} 
	\label{fig: 1000kt_1550_nm_peak_2}
\end{figure}

Since the maximum attenuation values in Fig. \ref{fig: max_atten_all} are independent of the attenuation of the clear air atmospheric channel ($A_{atm}$), the additional attenuation $A_{nuc}$ for the 1550~nm wavelength of the 1~Mt yield ensemble are plotted in Fig. \ref{fig: 1000kt_1550_nm_peak_2}. At 3~hours, 95\% of the 1~Mt simulations have no location where the debris contribution to attenuation is greater than 3~dB and the median 1~Mt simulation has a maximum attenuation of 0.9~dB. 

\begin{figure}[htp]
	\centering\includegraphics[width=.55\textwidth]{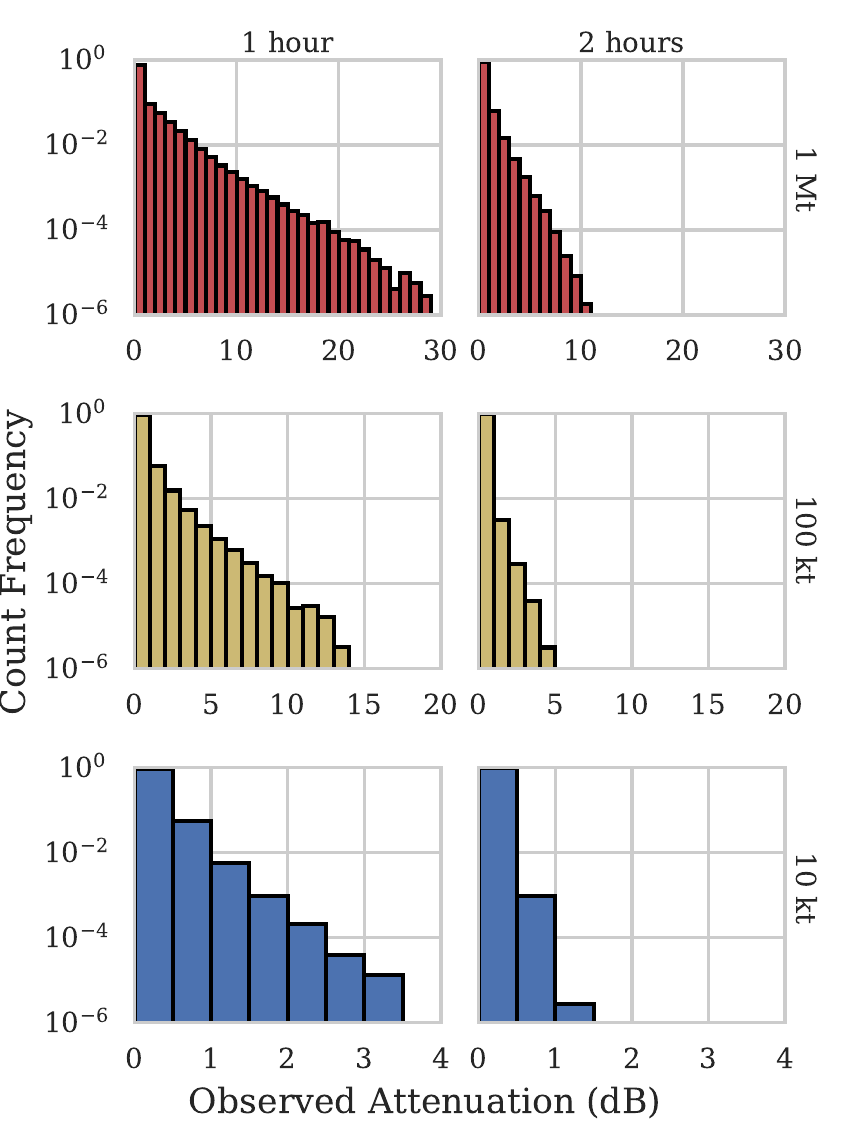}
	\caption{Attenuation of the 1550 nm wavelength at all nonzero grid locations for all simulations.} 
	\label{fig: attenuations}
\end{figure}

Where Fig.~\ref{fig: max_atten_all} illustrates the maximum observed attenuation in each simulation, Fig.~\ref{fig: attenuations} shows all observed nonzero attenuation results at all geographic grid locations for each ensemble. (Unlike the maximal histograms, the count frequency is given as a fraction of all observations in semi-log scaling to show the smaller bins. Note the very rapid reduction in count frequency from the first to second bin in all subplots.) In all cases and at all times, nearly all of the geographic grid locations experience less than 1~dB of attenuation due to nuclear debris. For the worst case of a 1~Mt yield and one hour after detonation, there is less than a 1\% chance that an affected location would experience added attenuation greater than 10 dB. By two hours, the likelihood of 10~dB attenuation is effectively zero. 

\subsection{Nuclear Link Loss - Secondary Smoke} \label{Smoke_Atten}

\begin{figure}[htp]
	\centering\includegraphics[width=.50\textwidth]{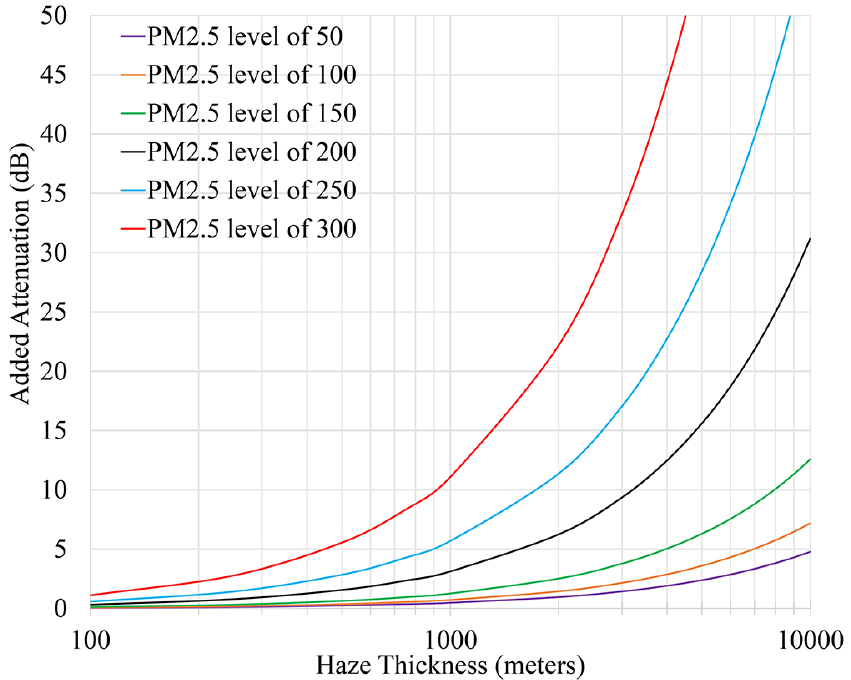}
	\caption{Added 1550 nm light attenuation as a function of haze thickness and its PM2.5 level.} 
	\label{fig:Haze}
\end{figure}

The last absorption/scattering mechanism that can cause degradation in the optical link is from elevated levels of particulate matter in the air from smoke and haze of a nearby burning city. In this scenario, we assume that a nuclear detonation has occurred in a nearby city, the debris of the nuclear cloud has passed, and that the smoke from the burning city is being blown in the direction of the ground station for the optical link. In smoke, the majority of particles are under 1~$\mu$m and so it is common to use the particulate matter 2.5 (PM2.5, typically with units of $\mu$g/m\textsuperscript{3}) equations to determine visibility in the smoke. The amount of smoke and haze from a burning city is difficult to quantify as each city has different parameters (such as fuel loading, city size, average wind speeds, etc.) though some quantification efforts have been performed~\cite{turco1990climate}.  A full treatment of smoke is beyond the scope of the present study. As a scoping study, however, a comparison of varying PM2.5 levels are considered  for an optical link.  The PM2.5 levels, shown in Fig.~\ref{fig:Haze} are used to calculate the visibility of the optical link as a function of haze thickness. The equation to relate visibility to PM2.5 levels is taken from~\cite{sun2020quantifying} and assumes relative humidity levels under 60~percent.

\begin{figure}[htp]
	\centering\includegraphics[width=.50\textwidth]{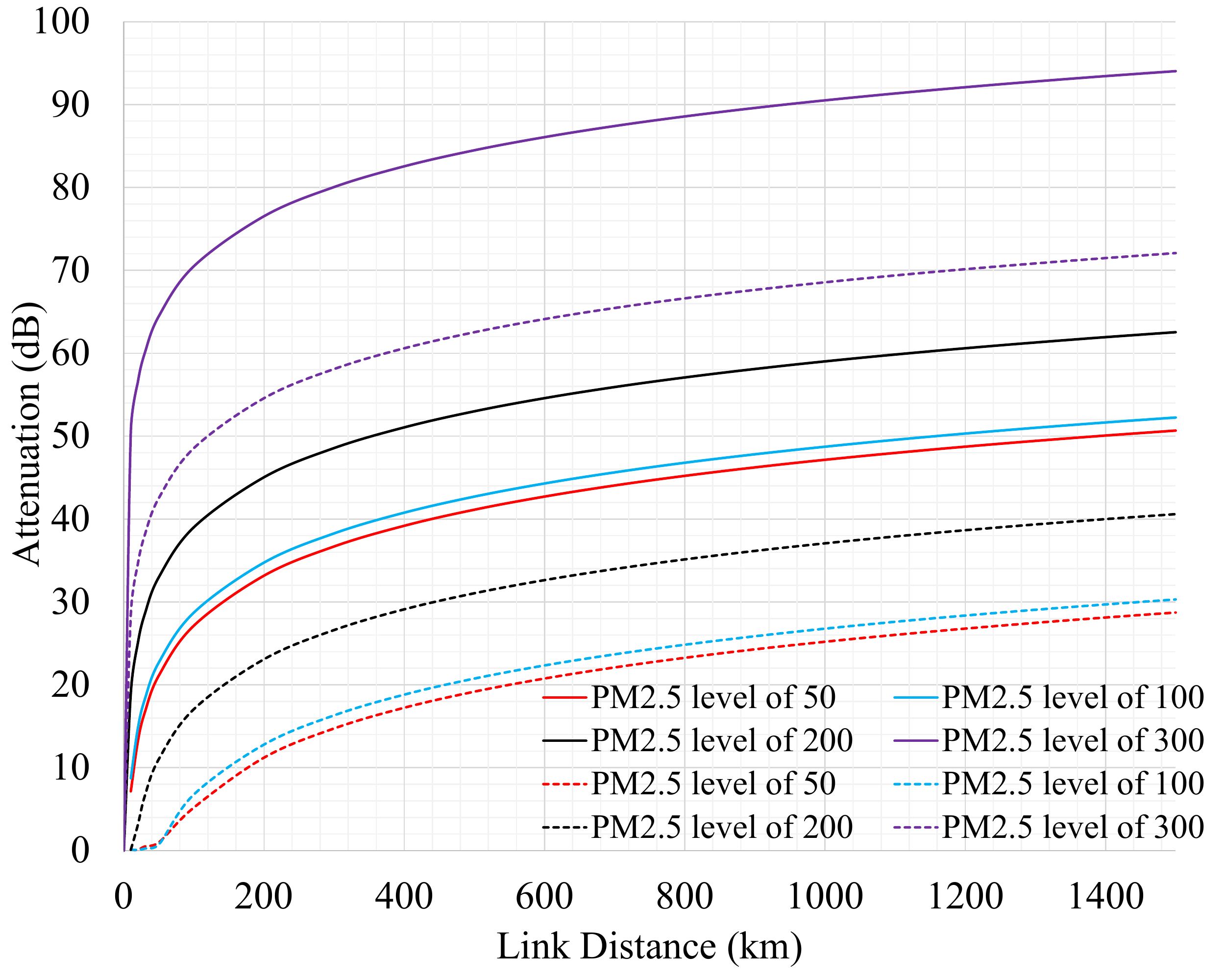}
	\caption{Optical link attenuation between a ground station and a satellite for both the downlink (solid line) and uplink (dashed line) scenarios, assuming the link propagates through 3 km of haze at an elevation angle of $90^{\circ}$} 
	\label{fig:Hazedb}
\end{figure}

The optical link attenuation calculation in Fig.~\ref{fig:Hazedb} assumes a 1550 nm uplink and downlink scenario between a ground station and a satellite; the satellite is assumed to have a 0.1~meter diameter optical aperture while the ground station is assumed to have a 1 meter diameter optical aperture. The amount of haze and smoke is set for a vertical thickness of 3 km for this study and the satellite is assumed to be directly overhead (elevation angle of $90^{\circ}$). Fig.~\ref{fig:Hazedb} is a best case scenario as the attenuation increases when the elevation angle is dropped from 90 degrees (attenuation increases as the inverse of the sine of the elevation angle).

\subsection{Nuclear Link Loss Summary} \label{Nuke_Summary}

Table \ref{tab: Summary} summarizes representative attenuation values for the nuclear cloud and debris terms in comparison to natural sources (clouds, city haze, and atmosphere) as well as the potential secondary effect of post-detonation fire. During the rise of the nuclear cloud up to stabilization, the attenuation of the signal through the cloud is several hundred~dB in addition to normal atmospheric losses. However, by an hour after detonation, atmospheric diffusion of the debris reduces the maximum attenuation to a median of 7~dB for a 1~Mt detonation, with 30~dB as the highest attenuation seen at 1~hour in any of the simulations. Attenuation from lower yields are substantially lower at 1~hour as well. 

\subsection{QSE}
So far, we have focused on computing the attenuation of an optical channel. Now we will describe how the attenuation translates into the degradation of transmitted heralded entanglement. To do so, we apply the formalism from Section \ref{QSE} to simulation data. The simulations assume that 1 million maximally entangled photon pairs are generated by sender for each measurement basis combination, nine million photon pairs total. The photon pair is in the polarization entangled Bell state
\begin{equation}
    \left| \Psi_{-} \right\rangle = \frac{1}{\sqrt{2}} \left( \left| H_{A}V_{B} \right\rangle - \left| H_{B}V_{A} \right\rangle \right).
\end{equation}
The sender detects one photon of the pair, heralding a single photon which is sent to the receiver. The transmission channel includes attenuation which has a chance of destroying the single photon. We vary the attenuation from 0~dB to 50~dB. From the collected data we estimate the state, the EOF from Eq.~\ref{EOF}, and the transmission efficiency estimate $\bar{\alpha}$, the attenuation is $1-\bar{\alpha}$. 

{\centering
\begin{minipage}[ct]{0.99\textwidth}
 {\centering
  \captionof{table}{Summary of Representative Signal Attenuation Terms (in dB).}
 \begin{tabular}{|p{0.25in}p{0.25in}p{0.25in}||p{0.25in}p{0.25in}p{0.25in}|p{0.25in}p{0.25in}p{0.25in}|p{0.25in}p{0.25in}p{0.25in}|}
 \hline
 \multicolumn{12}{|c|}{Nuclear Debris} \\
 \hline
 \multicolumn{3}{|c||}{Yield} & \multicolumn{3}{c|}{Nuclear Cloud$^a$} & \multicolumn{3}{c|}{1 Hour$^b$} & \multicolumn{3}{c|}{2 Hours$^b$} \\
 \hline
\multicolumn{3}{|c||}{1 Mt} & \multicolumn{3}{c|}{550} & \multicolumn{3}{c|}{30 (7)} & \multicolumn{3}{c|}{10 (2)} \\
 \hline
\multicolumn{3}{|c||}{100 kt} & \multicolumn{3}{c|}{350} & \multicolumn{3}{c|}{15 (2)} & \multicolumn{3}{c|}{5 (0.5)} \\
 \hline
\multicolumn{3}{|c||}{10 kt} & \multicolumn{3}{c|}{250} & \multicolumn{3}{c|}{3 (0.5)} & \multicolumn{3}{c|}{1 (0.1)} \\
 \hline
 \multicolumn{12}{|c|}{Clouds$^c$} \\
 \hline
 \multicolumn{4}{|c|}{Nimbostratus} & \multicolumn{4}{c|}{Cumulus} & \multicolumn{4}{c|}{Stratus} \\
 \hline
  \multicolumn{4}{|c|}{220} & \multicolumn{4}{c|}{70} & \multicolumn{4}{c|}{25} \\
 \hline
 \multicolumn{12}{|c|}{PM2.5 Pollutants} \\
 \hline
 \multicolumn{4}{|c|}{Fire Haze$^d$} &  \multicolumn{4}{c|}{Foreign City$^e$} & \multicolumn{4}{c|}{US City$^e$}   \\
 \hline
 \multicolumn{4}{|c|}{30} &  \multicolumn{4}{c|}{3} &  \multicolumn{4}{c|}{1} \\
 \hline
 \multicolumn{12}{|c|}{Atmosphere$^f$}  \\
 \hline  
 \multicolumn{6}{|c|}{Downlink} & \multicolumn{6}{c|}{Uplink} \\
 \hline
 \multicolumn{6}{|c|}{16.0} & \multicolumn{6}{c|}{34.0}        \\
 \hline
 \end{tabular} 
 \label{tab: Summary}\par }

\noindent\footnotesize{$^a$ Nuclear Cloud attenuation is calculated at the time of cloud stabilization (i.e., end of cloud rise): approx. 30~min \\ for 1~Mt, 10~min for 100~kt, and 5~min for 10~kt.}

\noindent\footnotesize{$^b$ Maximum attenuation for all simulations is given. The value in parentheses is the median maximum value across \\ all simulations.}

\noindent\footnotesize{$^c$ Cloud data taken from~\cite{5286364}}

\noindent\footnotesize{$^d$ Fire Haze data taken from~\cite{s20174796}}

\noindent\footnotesize{$^e$ Foreign City and US City pollution data taken from~\cite{ZHANG2020105862}}

\noindent\footnotesize{$^f$ \textit{L} = 400 km, \textit{D$_{S}$} = 0.1~m, \textit{D$_{GS}$} = 1~m.}
\end{minipage}
\newline
\newline
}

\begin{figure*}[htp]
	\centering\includegraphics[width=.9\textwidth]{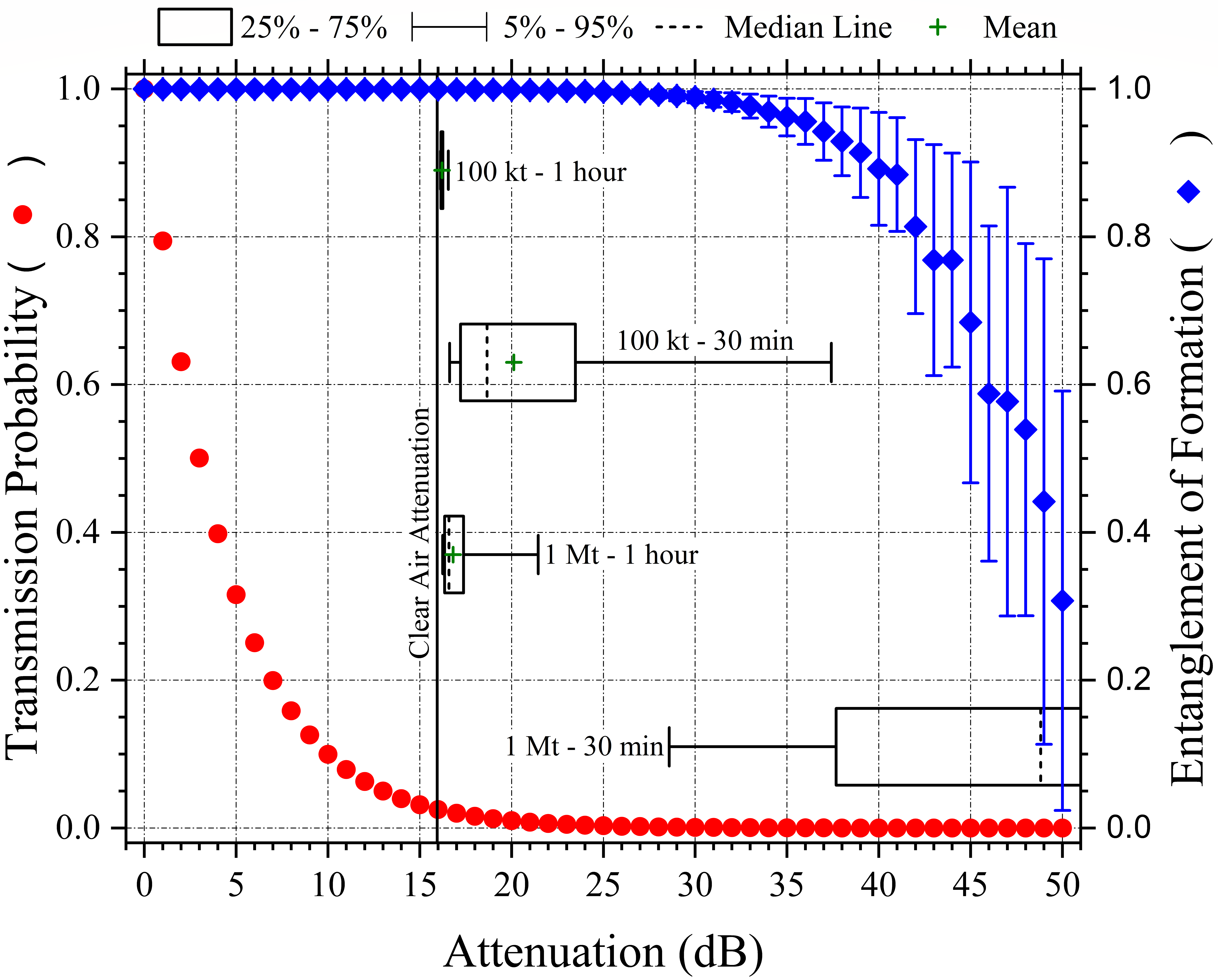}
	\caption{The entanglement estimate (with 1 standard deviation) and the transmission probability as a function of the loss in a nuclear disturbed environment after the detonation of a 1~Mt and 100~kt device. The solid black line is the downlink clear air attenuation calculated for a 400~km optical link between a space aperture diameter of 0.1~m and a ground station 1~m in diameter. Box plots are shown for 1~Mt and 100~kt explosions 30 minutes and 1 hour after detonation to show the evolution of the air and nuclear debris attenuation statistics with time. 10~kt data is not shown due to insignificant contribution to attenuation (see Table~\ref{tab: Debris + Air}).
} 
    \label{fig:Trans_Est}
\end{figure*}

We plot the entanglement and transmission estimates for each attenuation setting in Fig.~\ref{fig:Trans_Est}, using Eq.~\ref{Eq:dB_add}. The numerical data for the combined atmosphere/debris attenuation are given in Table~\ref{tab: Debris + Air}.

\begin{table*}
    \caption{Summary of the attenuation due to the nuclear debris and atmosphere for varying yields (kilotons) and times post-detonation (TPD) (hours). The clear air attenuation is assumed to be 16.0 dB for a downlink configuration optical link of 400 km between a 0.1 m diameter space aperture and a ground station with a 1 m aperture size. The mean, standard deviation (SD), median, and 95\% confidence intervals (CI) are shown for the nuclear debris attenuations as well as the total mean and median attenuations for the clear air and debris combined.}
    \centering
    \begin{tabular}{|c|c|ccccccc|}
    \hline
        Yield             & TPD & Nuclear & SD & Total & Nuclear & 95\% &  5\% & Total  \\
                          &     & Mean    &    & Mean  & Median  & CI   &  CI  & Median \\
         (kt)             & (h) &       \multicolumn{7}{c|}{(dB)}      \\
    \hline
    \multirow{3}{*}{1000} & 0.5 &  55.6 &   23.9 & 55.6 &   48.8 &  104. &   28.3 & 48.8 \\
                          & 1.0 &  9.12 &   5.07 & 16.8 &   7.65 &  20.0 &   3.58 & 16.6 \\
                          & 2.0 &  2.67 &   1.71 & 16.2 &   2.17 &  6.36 &  0.841 & 16.2 \\
    \hline
    \multirow{3}{*}{100}  & 0.5 &  18.0 &   9.16 & 20.1 &   15.3 &  37.4 &   7.93 & 18.7 \\
                          & 1.0 &  2.98 &   2.08 & 16.2 &   2.30 &  7.30 &  0.832 & 16.2 \\
                          & 2.0 & 0.688 &  0.589 & 16.1 &  0.496 &  1.87 &  0.145 & 16.1 \\
    \hline
    \multirow{3}{*}{10}   & 0.5 &  3.98 &   1.88 & 16.2 &   3.51 &  7.66 &   1.76 & 16.2 \\
                          & 1.0 & 0.598 &  0.411 & 16.1 &  0.481 &  1.40 &  0.147 & 16.1 \\
                          & 2.0 & 0.120 & 0.0970 & 16.1 & 0.0923 & 0.310 & 0.0473 & 16.1 \\
    \hline
    \end{tabular}
    \label{tab: Debris + Air}           
\end{table*}

As can be seen in Fig.~\ref{fig:Trans_Est}, our estimate for the transmission efficiency has almost no uncertainty due to the symmetry of the attenuation and the large dataset. More specifically, the sender gets one count for every photon pair generated. Whether or not the receiver receives a count is directly proportional to the attenuation. Thus, as simulated, this is a very low uncertainty measurement.

We also plot the entanglement of formation versus the loss in Fig.~\ref{fig:Trans_Est}. From this, one can discern the impact of a change in attenuation from a ``normal'' operating condition.  While our state never changes (it is still maximally entangled), the amount of data we generate decreases with increasing attenuation. Thus, the uncertainty in our entanglement estimates increases due to the finite photon statistics. One could always consider a longer integration time, but this illustrates that there are fundamental limits to how much attenuation is allowable for practical quantum applications.  In addition, as satellite platforms may be moving and hence not always aligned, opportunities to collect data may not have a high duty cycle further limiting the data set size.

Application of this approach to a dynamic attenuation is done by analyzing smaller chunks of data from discrete time bins. Analysis of estimates from short time intervals in ``real-time'' allows tracking channel dynamics.

The spread of attenuation values in Fig.~\ref{fig:Trans_Est} are calculated using Eq.~\ref{Eq:dB_add} assuming that $A_{atm}$~=~16~dB and $A_{nuc}$ is the nuclear debris values, obtained from the DELFIC simulation and presented in Table~\ref{tab: Debris + Air}. The spread of values is meant to highlight the degradation of the optical channel as a function of time due to the debris when compared to the normal operating conditions. The mean and the median are provided to show the skew of the distribution. The debris only impairs the optical channel during the first hour, with the severity depending on the explosive yield. Despite the adverse effect on optical channel, only the 1~Mt explosion is able to produce enough attenuation in the first hour where the entanglement severely suffers.

\section{Conclusions}

The research of this paper addresses the potential for airborne nuclear debris to interfere with optical or quantum communications between a ground/satellite transmitter/receiver system. If a nuclear weapon were detonated near the surface of the earth (where a significant amount of ground soil is entrained into the fireball), the rising cloud may contain up to millions of tons of debris. During the cloud rise, much of the debris remains contained in the cloud due to the internal vorticity. If an optical or quantum signal were transmitted through the rising cloud, the signal loss would be in the hundreds of dB, likely rendering successful communication impossible. This condition would last until the cloud reaches its peak altitude, at which point the internal vorticity is completely dissipated and the debris transports in the atmosphere according to wind and gravitational forces. The dissipation of the debris rapidly decreases the peak attenuation of the signal. By the time an hour has passed after the detonation, the maximum simulated attenuation of the signal is bounded within 30~dB for a 1~Mt surface detonation. After two~hours, the peak attenuation is reduced to about 10 to 11~dB for 1~Mt. These peak values, however, are the worst cases in all scenarios, where the signal is being transmitted through the location of greatest loss. For the rising cloud, this is a cross-sectional radius of a few km that is generally found relatively close to the ground zero of the detonation. By one or two~hours, the point of peak attenuation may have moved to 100~km or more downwind of ground zero, but the area affected by this attenuation is still relatively small, and the rapid speed of upper air winds will generally ensure that communications at a fixed point are interrupted for only a small window of time.

When looking at the entire region where airborne debris has a measurable attenuation of the signal, it is very unlikely that a ground/satellite system within this region would experience significant attenuation. Even for a 1~Mt explosion at 1~hour, about 90\% of all the affected area experience 1~dB or less of attenuation, and 99\% of all the area experience less than 1~dB attenuation. If the non-nuclear atmospheric loss is assumed to be about 20~dB, then the additional attenuation burden of the nuclear debris is a very small fraction of the total signal loss. Whether this additional burden is significant will be a function of the transmitter/receiver system.

In contrast, the smoke from potential fires caused by a near-surface nuclear detonation may provide a larger, more persistent source of attenuation downwind of the detonation site. The PM2.5 pollutants may cause losses on the order of 30~dB throughout the affected region, and if these fires persist over hours or days, then the attenuation would potentially be present for a much longer time. The general concerns are thus (in chronological order):

\begin{enumerate}
    \item The rising nuclear cloud, if the transmitter/receiver system is close to, and downwind of, the detonation, but not so close as to be affected by the initial blast or other localized effects. 
    \item The dispersed nuclear debris, up to an hour after detonation for very large yields, and for less time for smaller yields.
    \item Smoke and soot from fires caused by the detonation, which may persist for hours or days after the detonation and become a source of large attenuation for the affected regions. Given that the winds will not be constant, the region of affected transmission may change over time.
\end{enumerate}

Our analyses focused on 1550~nm wavelengths due to their relatively low attenuation through normal atmospheric conditions. A brief inspection of the 800~nm wavelength revealed that the attenuation through debris -- whether the rising cloud, the transported debris, or the smoke -- was generally insensitive to the difference between these wavelengths due to the broad distribution of particle sizes (typically from 1 to 100~~$\mu$m). 

Lastly, we provide estimates on the impact to entanglement transmission through a nuclear-disturbed optical channel. By adding the effects of the nuclear debris (in dB) to that of the atmosphere, an estimate of the potential degradation of entanglement is found for various yields and times after detonation. In the case of an actual need to transmit a quantum signal through a nuclear disturbed environment, these loss estimates may be applied to the particular transmitter/receiver system, atmospheric conditions, and nuclear-induced disturbance to assist in operational decision-making. By maintaining the assumption of independence of the atmospheric losses and the nuclear losses, the potential attenuation and signal degradation may be rapidly estimated by adding the separate loss terms in dB. As our research is intended to provide upper bound estimates on the potential losses, we minimize the possibility of predicting `false positive' transmission rates using this method.

This research focused on quantifying the severity of potential transmission attenuation due to airborne nuclear debris, in terms of total attenuation and time of significant attenuation. While the analysis is applicable to both conventional and quantum signals, analysis of measures to counteract these effects are beyond the scope of the research, but merit some discussion. As noted, this work assumed communications between a single satellite and a single ground station. Given the localized effect of nuclear debris attenuation, it is entirely possible that an optical signal or entangled photon may simply be sent to a different receiver, or sent from a different transmitter if such a transmission is time-critical. This decision is best analyzed by the operator of the communications network; in our opinion, however, this course of action is likely to be viable at any time after detonation. Alternatively, changing to a modality (e.g., radio frequency communication, for classical-only communications) that is insensitive to the presence of debris may provide a solution. For solely classical communications, the use of high-intensity lasers to mitigate turbulence~\cite{salame2007} or fog~\cite{Schimmel2018} may be considered. Similar to a change in wavelength, this obviates the use of single-photon communications, which may or may not be desirable depending on the needs of the user. 

\appendix
\section{Parametrizing the two-qubit density matrix}\label{densityMatrixAppendix}

\renewcommand{\theequation}{A.\arabic{equation}}

 We briefly describe the parametrization of the two-qubit density matrix here, for a complete description of this topic see~\cite{williams2017quantum}. For a $4$-dimensional Hilbert space, the density matrix is formed using the Cholesky decomposition, requiring $\rho=L(\!\tau\!)L(\!\tau\!)^\dagger$ with
\begin{equation}L(\!\tau\!)\! =\!\!\left(\!\begin{array}{ccccc}
    L_{11} (\!\tau\!)\!\!& 0& 0 & 0 \\
    L_{21}(\!\tau\!)\! \! & L_{22}(\!\tau\!)\!\! & 0 & 0  \\
    L_{31}(\!\tau\!)\! \! & L_{32}(\!\tau\!)\!\!& L_{33}(\!\tau\!)\!\! &  0 \\
    L_{41}(\!\tau\!)\! \! & L_{42}(\!\tau\!)\!\!& L_{43}(\!\tau\!)\!\! &  L_{44}(\!\tau\!) 
    \end{array}\right)\ ,
    \vspace{5pt}
    \end{equation}
which is a lower triangular matrix with positive real diagonal elements. The parameter set $\tau$ includes $15$ parameters which describe a unique density matrix.

The elements $L_{ij}$ may be written as
\begin{equation}
\begin{aligned} L_{ij} (\!\tau\!)&=U_i V_{ij}\phantom{0}\quad\quad (j\leq i)\ , ~\text{and} \\
L_{ij} (\!\tau\!)&=0\phantom{U_i V_{ij}}\quad\quad (j> i)\ , \end{aligned}
\end{equation}
where
\begin{equation}
\begin{aligned}U_{1}&=\cos\left(u_1\right), \\
U_{k}&=\cos\left(u_k\right)\prod_{j=1}^{k-1}\sin\left(u_j\right)\quad (1<k<n)\ , \\
U_{n}&=\sin\left(u_n\right), \\
V_{ii}&=1, \\
V_{i1}&=\cos\left(\theta_{i1}\right)e^{i\phi_{i1}}\quad (i>1)\ ,~\text{and}  \\
V_{ik}&=\cos\left(\theta_{ik}\right)e^{i\phi_{ik}}\prod_{j=1}^{k-1}\sin\left(\theta_{ij}\right)\quad (1<k<i). 
\end{aligned}
\end{equation}

Our simulation has dimension $n$$=$$4$, and the parameterized matrix elements of $L(\!\tau\!)$ are
\begin{equation}
\begin{aligned}L_{11}(\!\tau\!)&\!=\!\cos(u_1)\ ,\\
    L_{21}(\!\tau\!)&\!=\!\sin(u_1)\cos(u_2)\cos(\theta_{21})e^{i \phi_{21}},\\
    L_{22}(\!\tau\!)&\!=\!\sin(u_1)\cos(u_2) \sin(\theta_{21})\ ,\\
        L_{31}(\!\tau\!)&\!=\!\sin(u_1)\sin(u_2)\cos(u_3)\cos(\theta_{31})e^{i \phi_{31}},\\
L_{32}(\!\tau\!)&\!=\!\sin(u_1)\sin(u_2)\cos(u_3)\sin(\theta_{31})\cos(\theta_{32})e^{i \phi_{32}},\\
L_{33}(\!\tau\!)&\!=\!\sin(u_1)\sin(u_2)\cos(u_3)\sin(\theta_{31})\sin(\theta_{32})\ ,\\
        L_{41}(\!\tau\!)&\!=\!\sin(u_1)\sin(u_2)\sin(u_3)\cos(\theta_{41})e^{i \phi_{41}},\\
        L_{42}(\!\tau\!)&\!=\!\sin(u_1)\sin(u_2)\sin(u_3)\sin(\theta_{41})\cos(\theta_{42})e^{i \phi_{42}},\\
        L_{43}(\!\tau\!)&\!=\sin(u_1)\sin(u_2)\sin(u_3)\sin(\theta_{41})\sin(\theta_{42})\cos(\theta_{43})e^{i \phi_{43}},~\text{and}\\
        L_{44}(\!\tau\!)&\!=\!\sin(u_1)\sin(u_2)\sin(u_3)\sin(\theta_{41})\sin(\theta_{42})\sin(\theta_{43})
\end{aligned}
\end{equation}
with $u_{i}\in[0,\frac{\pi}{2}]$, $\theta_{ij}\in[0,\frac{\pi}{2}]$, and $\phi_{ij}\in[0,2\pi]$. 
The parameter set is then explicitly \begin{equation}\tau=\left\{u_1,u_2,u_3,\theta_{21},\theta_{31},\theta_{32},\theta_{41},\theta_{42},\theta_{43},\phi_{21},\phi_{31},\phi_{32},\phi_{41},\phi_{42},\phi_{43}\right\}\textrm{.}\end{equation}

\section*{Acknowledgments}
This work was supported by Defense Threat Reduction Agency Award HDTRA1-93-1-201.

This manuscript has been co-authored by UT-Battelle, LLC, under contract DE-AC05-00OR22725 with the US Department of Energy (DOE). The US government retains and the publisher, by accepting the article for publication, acknowledges that the US government retains a nonexclusive, paid-up, irrevocable, worldwide license to publish or reproduce the published form of this manuscript, or allow others to do so, for US government purposes. DOE will provide public access to these results of federally sponsored research in accordance with the DOE Public Access Plan (http://energy.gov/downloads/doe-public-access-plan).

\section*{Disclosures}
The authors declare no conflicts of interest.



\bibliographystyle{apsrev4-1}
\bibliography{citations.bib}


\end{document}